\documentclass[aps,orb,twocolumn,amssymb,showpacs,prd]{revtex4-1}
\usepackage{graphicx}
\usepackage{color}
\begin{document}

\def\b{{\beta }}
\def\z{{\zeta }}
\def\be{\begin{equation}}
\def\ee#1{\label{#1}\end{equation}}
\def\d{\textsf{d} }
\def\e{\textsf{e} }
\def\x{\textsf{x} }
 \def\bx{\mathbf{x} }
 \def\bp{\mathbf{p} }
 \def\p{\textsf{p} }
  \def\n{\textsf{n} }
 \def\pp{\textsf{P} }
 \def\s{\textsf{s} }
 \def\no{\nonumber}
\def\lb{\label}
\def\c{\textsf{c} }
\def\f{\textsf{f} }
\def\g{\textsf{g} }
\def\x{\textsf{x} }
\def\G{\textsf{G}}
\def\M{\textsf{M}}
\def\D{\textsf{D}}

\newcommand{\ben}{\begin{eqnarray}}
\newcommand{\een}{\end{eqnarray}}
\newcommand{\w}{{\bf **}}
\newcommand{\anf}{``}

 %%%%%%%%%%%%%%%%%%%%%%%%%%%%%%%%%%%%%%%%%%%%%%%%%%%%%%%%%%%%%%%%%%%
\title{Temperature oscillations of a gas in circular geodesic motion in the Schwarzschild field}
\author{Winfried Zimdahl}\email{winfried.zimdahl@pq.cnpq.br}
\affiliation{Departamento de F\'{\i}sica, Universidade Federal do Esp\'{\i}rito Santo, 29075-910 Vit\'oria, Brazil}
\author{Gilberto M. Kremer}\email{kremer@fisica.ufpr.br}
\affiliation{Departamento de F\'{\i}sica, Universidade Federal do
Paran\'a, 81531-980 Curitiba, Brazil}

\begin{abstract}
We investigate a Boltzmann gas at equilibrium with its center of mass moving on a circular geodesic
in the Schwarzschild field. As a consequence of Tolman's law we find that a central comoving observer measures oscillations of the temperature and of other thermodynamic quantities with twice the frequencies that are known from test-particle motion. We apply this scheme to the gas dynamics in the gravitational fields of the planets of the Solar System as well as to strong-field configurations of neutron stars and black holes.
\end{abstract}
 \maketitle

 %%%%%%%%%%%%%%%%%%%%%%%%%%%%%%%%%%%%%%%%%%%%%%%%%%%%%%%%%%%%%%%%%%%
\section{Introduction}

Relativistic gas theory started as early as 1911 with the work of J\"{u}ttner \cite{Jue} who derived the one-particle equilibrium distribution function for a Boltzmann gas.
The covariant form of Boltzmann's equation was obtained by Lichnerowicz and Marrot \cite{LiMa}. Equilibrium conditions for a relativistic gas were shown to imply the existence of a timelike Killing vector \cite{Cher}. One of the consequences is Tolman's law \cite{To1,To2} and another one is Klein's law \cite{Klein}.

General-relativistic gaseous fluids are of interest in cosmology and astrophysics. In many occasions the material content of the Universe is modeled in fluid dynamical terms. Then it is the energy-momentum tensor of the fluid which appears on the right-hand side of Einstein's equations and determines the gravitational field. In an astrophysical context it is also the behavior of a gas on a given gravitational background which is important. The probably most interesting case here is the accretion of (gaseous) matter towards compact objects like black holes.
As an example we mention the observed high-frequency quasiperiodic oscillations in the fluxes from x-ray binaries which are supposed to be of general-relativistic origin \cite{Abramowicz,Klis,Lamb,Claudio}.

Any gas-dynamical analysis starts by establishing the relevant equilibrium configurations of the system under consideration. For a Boltzmann gas one distinguishes local equilibrium from global equilibrium.  The former relies on the mere equilibrium structure of the one-particle distribution function; the latter adds conditions on the quantities that appear in this structure. One of these conditions implies the existence of a timelike Killing vector, equivalent to a stationarity condition for the metric. Generally, a global equilibrium is therefore incompatible with an expanding Universe, while local equilibrium states of the cosmic substratum are suitable starting points in standard cosmology.

The present paper is devoted to a global equilibrium configuration on the background of the static Schwarzschild metric.
We consider a Boltzmann gas with its center of mass moving on a circular geodesic of this metric and
we study the equilibrium thermodynamics of this system as seen by a comoving observer on the geodesic.
While this is clearly an idealized situation, we hope that it may capture some typical features and serve as a starting point for more realistic problems in astrophysics where either the equilibrium hypothesis or the circular geodesic nature of the worldline or both are generalized.

The lowest-order gravitational effects that a freely falling observer can detect locally are conveniently described with the help of Fermi normal coordinates. These coordinates are Minkowskian on the geodesic while gravitation at lowest order manifests itself in quadratic corrections in the spacelike geodesic distance, orthogonal to the observer's timelike trajectory.
We apply a description of this type to the gas motion, admitting additionally a pure spatial rotation. This corresponds to using a ``proper reference" frame (cf. Ref.~\cite{MTW}) up to second order.
Fixing this frame determines, via Tolman's law \cite{To1,To2}, the temperature profile of the gas as measured by a central, geodesic observer.
It turns out that the comoving observer measures oscillations of the temperature and other thermodynamic quantities with frequencies that are double the frequencies known from test-particle motion in the Schwarzschild field.

The structure of the paper is as follows. In Sec.~\ref{Boltzmann} we recall basic properties of the Boltzmann equation for the one-particle distribution function and its equilibrium solution. Section \ref{Equilibrium} is devoted to the corresponding equilibrium conditions. On this basis Tolman's and Klein's laws are derived in Sec.~\ref{TolmanKlein}.
Section \ref{Schwarzschild} reviews the circular test-particle motion in the Schwarzschild field.
Fermi normal tetrads are introduced and discussed in Sec.~\ref{Fermi} on the basis of which the temperature profile
is obtained in Sec.~\ref{profile}. With the help of the equation of geodesic deviation we reproduce, in Sec.~\ref{geodesic},  the basic linear system which describes oscillatory motions around the central geodesic. The corresponding oscillations of the temperature and the other thermodynamic quantities are discussed  in  Sec.~\ref{thermodynamic}.
Section~\ref{summary} summarizes our results.

 \section{Boltzmann equation and equilibrium distribution function}
 \label{Boltzmann}

A relativistic gas of particles with rest mass $m$ in a spacetime with metric tensor $g_{\mu\nu}$  is characterized by the spacetime coordinates $(x^\mu)=(ct, \bx)$ and the momenta  $(p^\mu)=(p^0, \bp)$. Because of the mass shell condition
$g_{\mu\nu}p^\mu p^\nu=-m^2c^2$ a state of the gas is described by the one-particle distribution function $f(\bx, \bp, t)$ in the phase space spanned by $x^\mu$ and $\bp$. The distribution function is defined in such a way that $f(\bx, \bp, t)d^3x d^3 p$ is the number of particles in the volume element $d^3 x\,d^3 p$ at the time $t$.

The  evolution of the distribution function $f(\bx, \bp, t)$ in the phase space is governed by the Boltzmann equation (see e.g., Ref.~\cite{CKr})
\ben\lb{1}
p^\mu\frac{\partial f}{\partial x^\mu}-\Gamma_{\mu\nu}^\sigma p^\mu p^\nu
\frac{\partial f}{\partial p^\sigma}=Q(f,f),
\een
where the $\Gamma_{\mu\nu}^\sigma$ are the Christoffel symbols and $Q(f,f)$ is the collision operator of the Boltzmann equation. For an equilibrium distribution $f^{(0)}$ the collision operator $Q(f^{(0)},f^{(0)})$ vanishes and, for a classical gas, $f^{(0)}$ becomes the Maxwell-J\"uttner distribution function
\ben\lb{2}
f^{(0)}=\frac{\n}{4\pi k T m^2 c  K_2(\z)}\exp\left({-\frac{U^\tau p_\tau}{k T}}\right).
\een
Here, $\n$ is the particle number density, $T$ is the temperature, $U^\tau$ is the four-velocity (with $U^\mu U_\mu=-c^2$) and $k$ is the Boltzmann constant.
Furthermore,
\be
K_n(\zeta)=\left(\frac{\zeta}{2}\right)^n\frac{\Gamma(1/2)}{\Gamma(n+1/2)}\int_{1}^\infty e^{-\zeta y}\left(y^2-1\right)^{n-1/2}\,dy
\ee{3}
denotes modified Bessel functions of the second kind with $\z=mc^2/kT$. The parameter $\z$ is the ratio of the rest energy $mc^2$ of a gas particle and the thermal energy $kT$ of the gas. The nonrelativistic limit corresponds to $\z\gg1$, while the ultrarelativistic limit is obtained for $\z\ll1$.

With the help of the equilibrium distribution function (\ref{2}) we may calculate the energy-momentum tensor
\begin{equation}\label{4a}
T^{\mu\nu}=c\int p^\mu p^\nu f^{(0)}\sqrt{-g}\frac{d^3p}{p_0}=\frac{\e \,\n}{c^2}U^\mu U^\nu +\p g^{\mu\nu}\,,
\end{equation}
where $\e$ is the energy per particle and $\p$ is the pressure:
\begin{equation}\label{5a}
\e=mc^2\left[\frac{K_3(\z)}{K_2(\z)}-\frac{1}{\z}\right],\qquad \p=\n kT .
\end{equation}
Likewise we obtain
the entropy-flow vector
\begin{equation}\label{4b}
S^\mu=-k\,c\int p^\mu f^{(0)}\ln f^{(0)} \sqrt{-g}\frac{d^3p}{p_0}=\n\,\s\, U^\mu
\end{equation}
with the entropy per particle
\begin{equation}\label{5b}
\s=k\left\{\ln\left(\frac{4\pi kTm^2cK_2(\z)}{e\n}\right)
+\z\frac{K_3(\z)}{K_2(\z)}\right\}.
\end{equation}
The Gibbs function per particle $\g=\e-T\s+\p/\n$ may be  identified with the chemical potential $\mu$:
\ben\lb{6a}
\mu\equiv\g=kT\left[\ln\left(\frac{e\n}{4\pi kTm^2cK_2(\z)}\right)\right].
\een
In terms of the chemical potential the equilibrium distribution function (\ref{2}) is written as
\ben\lb{7}
f^{(0)}=\exp\left({\frac{\mu}{kT}-1-\frac{U^\tau p_\tau}{k T}}\right).
\een

\section{Equilibrium conditions}
\label{Equilibrium}

While the right-hand side of Boltzmann's equation (\ref{1}) vanishes identically for $f = f^{(0)}$,
its left-hand side imposes restrictions on the
fields that appear in this distribution function. Indeed, inserting (\ref{7})
into (\ref{1}) one finds the polynomial equation
\ben\lb{8}
p^\nu\partial_\nu\left[{\mu\over kT}\right]-{1\over 2}p^\mu p^\nu\left\{\left[
{U_\mu\over kT}\right]_{;\nu}+\left[{U_\nu\over kT}\right]_{;\mu}\right\}=0.
\een
Since this equation is valid for all values of $p^\mu$ it implies
\ben\lb{9}
\partial_\nu\left[{\mu\over kT}\right]=0,\qquad
\left[
{U_\mu\over kT}\right]_{;\nu}+\left[{U_\nu\over kT}\right]_{;\mu}=0,
\een
provided the  particles have nonvanishing rest mass. The 
right-hand expression of
Eq.~(\ref{9}) is the so-called  Killing equation
and ${U_\nu\over kT}$ is a (timelike) Killing vector.
The Killing equation can be rewritten as
\ben\lb{9a}
U_{\mu;\nu}+U_{\nu;\mu}-\frac1{T}\left(T_{,\nu}U_\mu+T_{,\mu}U_\nu\right)=0.
\een
By suitable projections proportional and perpendicular to $U^{\alpha}$ we get the relations
\ben\lb{10a}
 \dot T=0,\quad \mathrm{and} \quad \dot U_\mu+\frac{c^2}{T} \nabla_{\mu}T=0,
\een
respectively. Here
\ben\lb{10b}
\dot T\equiv U^\mu\partial_\mu T,\quad\dot U_\mu\equiv U^\nu{U_\mu;\nu},\quad \nabla_{\mu}T \equiv h^{\nu}_{\mu}T_{,\nu},
\een
where $h_{\mu\nu} = g_{\mu\nu}+c^{-2}U_{\mu}U_{\nu}$.
Equations (\ref{10a}) may be interpreted as follows: in equilibrium a gas must have a stationary temperature and its acceleration must be counterbalanced by a spatial temperature gradient. In general, the
right-hand expression of condition (\ref{10a}) is therefore not compatible with a geodesic fluid motion which would require $\dot U^\mu=0$. We shall return to this point in Sec. \ref{geodesic}.
The trace of (\ref{9a}) yields ${U^{\mu}}_{;\mu}=0$; i.e., equilibrium requires a vanishing expansion scalar.

\section{Tolman and Klein laws}
\label{TolmanKlein}

In order to derive Tolman's law \cite{To1,To2} let us consider a fluid at rest such that
\ben\lb{11a}
(U^\mu)=\left(\frac{c}{\sqrt{-g_{00}}},\bf0\right).
\een
Taking into account that the existence of a timelike Killing vector amounts to a stationary metric, the acceleration term becomes
\ben\lb{11b}
\dot U^\mu=-\frac{c^2}{g_{00}}\Gamma_{00}^\mu=\frac{c^2}{2g_{00}}g^{\mu\nu}g_{00,\nu}.
\een
From the right-hand expression of (\ref{10a}) and from (\ref{11b}) we have
\ben\lb{12}
c^2g^{\mu\nu}\left[\ln\left(\sqrt{-g_{00}}\,T\right)\right]_{,\nu}=0,
\een
which implies Tolman's law
\begin{equation}\label{Tolman}
 \sqrt{-g_{00}}\,T = \mathrm{constant}\,.
\end{equation}

Klein's law follows from Tolman's law and the first of the equilibrium conditions in (\ref{9}) and reads $\sqrt{-g_{00}}\,\mu$=constant. It was obtained here from the equilibrium conditions applied to the Maxwell-J\"uttner distribution, but it can also be derived on purely thermodynamical grounds as in Klein's original
paper \cite{Klein}.

\section{Schwarzschild metric}
\label{Schwarzschild}

The Schwarzschild metric is
\ben\no
 ds^2=-{\left(1-\frac{2\mathcal{M}}{ r}\right)}\left(dx^0\right)^2+\frac{1}{\left(1-\frac{2\mathcal{M}}{ r}\right)}dr^2
 \\\lb{13}
 +r^2\left(d\vartheta^2+\sin^2\vartheta d\varphi^2\right)=-c^2d\tau^2,
 \een
where $\mathcal{M}=GM/c^2$ and $G$ denotes the gravitational constant.

The Lagrangian of the orbital motion  of a test particle with rest mass $m$ in the plane $\vartheta=\pi/2$ is
\ben\no
\mathcal{L}=\frac{m}2\left[\frac1{1-\frac{2\mathcal{M}}{ r}}\left(\frac{dr}{d\tau}\right)^2+r^2\left(\frac{d\varphi}{d\tau}\right)^2\right.
\\\lb{14}\left.
-\left(1-\frac{2\mathcal{M}}{ r}\right)\left(\frac{dx^0}{d\tau}\right)^2\right].
\een
From this expression we infer that $\varphi$ and $x^0$ are cyclic coordinates so that the corresponding generalized momenta $p_\varphi$ and $p_0$ are conserved and read
\ben\lb{15a}
p_\varphi=\frac{\partial \mathcal{L}}{\partial(d\varphi/d\tau)}=mr^2\frac{d\varphi}{d\tau}=l,\\\lb{15b}
p_0=\frac{\partial \mathcal{L}}{\partial(dx^0/d\tau)}=-m\left(1-\frac{2\mathcal{M}}{ r}\right)\frac{dx^0}{d\tau}=
-\frac{E}{c}.
\een
Here, $l$ and $E$ denote the angular momentum and the energy of the particle, respectively.

A circular orbit (constant radius $r$) is characterized by
\ben\lb{16}
\widetilde E^2=\left(1-\frac{2\mathcal{M}}{ r}\right)\left[1+\frac{\widetilde l^2}{r^2}\right],
\een
thanks to (\ref{13}), (\ref{15a}) and (\ref{15b}). Here we have introduced the energy in units of $mc^{2}$, $\widetilde E\equiv E/mc^2$ and the angular momentum in units of $mc$, $\widetilde l\equiv l/mc$.

The motion of the test particle is obtained from  (\ref{13}), (\ref{15a}) and (\ref{15b}) and can be written as
(see, e.g., \cite{MTW})
\ben\lb{16a}
\left(\frac{dr}{dc\tau}\right)^2+\widetilde V^2=\widetilde E^2,
\een
where $\widetilde V^2$ denotes the effective potential
\ben\lb{16b}
\widetilde V^2=\left(1-\frac{2\mathcal{M}}{ r}\right)\left[1+\frac{\widetilde l^2}{r^2}\right].
\een
The possible circular orbits are found by searching the extreme values of the effective potential $\widetilde V^2$ which results in
\ben\lb{17}
\widetilde l^2=\frac{\mathcal{M}r^2}{r-3\mathcal{M}}.
\een

Hence, a test particle of rest mass $m$ in orbital motion with $\vartheta=\pi/2$ is characterized by constant values  of angular momentum (\ref{17}) and energy
\ben
\widetilde E^2=\left(1-\frac{2\mathcal{M}}{ r}\right)^2\frac{r}{r-3\mathcal{M}},
\een
thanks to (\ref{16}) and (\ref{17}).

Since $d\varphi/d\tau$ is given by (\ref{15a}) and the angular momentum by (\ref{17})
it follows by integration that
\ben
 \varphi=\frac1{r}\sqrt{\frac{\mathcal{M}}{r-3\mathcal{M}}}\,\tau
\een
for a circular orbit. The corresponding angular
frequency of the particle motion is
\begin{equation}\label{ophi}
\omega_{\varphi} = \omega_{N}\sqrt{\frac{r}{r-3\mathcal{M}}},\qquad
\omega_{N}  = \sqrt{\frac{GM}{r^{3}}} ,
\end{equation}
where $\omega_{N}$ is the Newtonian frequency in the limit for $r\gg \mathcal{M}$.

If the particle is slightly displaced from the exact circular motion in radial direction, there exists another oscillation frequency which is given by half the second derivative of (\ref{16b}) combined with (\ref{17}) which reads \cite{Wald}
\begin{equation}\label{or}
\omega_{r} = \omega_{N}\sqrt{\frac{r-6\mathcal{M}}{r-3\mathcal{M}}}.
\end{equation}
Obviously, both frequencies only coincide for $r\gg \mathcal{M}$. Their difference gives rise to a precession of the particle motion within the orbital plane.

The nonvanishing components of the curvature tensor for $\vartheta=\pi/2$ are:
\ben\lb{18a}
R_{\bar0\bar1\bar0\bar1}=-\frac{2\mathcal{M}}{r^3},\qquad
R_{\bar0\bar2\bar0\bar2}=\frac{\mathcal{M}}{r}
\left(1-\frac{2\mathcal{M}}{ r}\right),\\\lb{18b}
R_{\bar1\bar2\bar1\bar2}=-\frac{\mathcal{M}}{r}\frac1{1-\frac{2\mathcal{M}}{ r}},\qquad
R_{\bar2\bar3\bar2\bar3}=2\mathcal{M}r,
\\\lb{18c}
R_{\bar0\bar3\bar0\bar3}=R_{\bar0\bar2\bar0\bar2},\qquad R_{\bar1\bar2\bar1\bar2}=R_{\bar1\bar3\bar1\bar3}.
\een
Here, the overbar denotes the original Schwarzschild coordinates according to (\ref{13}).

\section{Fermi normal coordinates}
\label{Fermi}

Now we consider a rarefied gas, say, inside a spacecraft, in a circular orbit around an object with mass $M$.
The contributions of the spacecraft and the gas to the gravitational field are assumed to be negligible.
Let the center of mass of the gas move on a circular geodesic of the Schwarzschild metric.
An observer at the center of mass will conveniently use Fermi normal coordinates to describe local gravitational effects in the vicinity of the geodesic \cite{MM}. These are comoving, time-orthogonal coordinates with the center of mass at rest in the origin. The time coordinate is the proper time $\tau$ of the center on the geodesic.
The spatial coordinates are orthogonal spacelike geodesics parametrized by the proper distance.

Quite generally, the components of the metric tensor up to the second order in the deviations from the geodesic
in Fermi normal coordinates are \cite{MM,MTW}
\ben\lb{19a}
&&g_{\hat0\hat0}=-1-R_{\hat0\hat n\hat0\hat m}x^{\hat n}x^{\hat m},\\\lb{19b}
&&g_{\hat0\hat i}=-\frac23 R_{\hat0\hat n \hat{i}\hat m}x^{\hat n}x^{\hat m},\\\lb{19c}
&&g_{\hat i\hat j}=\delta_{ij}-\frac13 R_{\hat i\hat n\hat j\hat m}x^{\hat n}x^{\hat m},
%\\&&\vert g\vert=\det(g_{\mu\nu})=1+\frac13 \left(R_{\hat n\hat m}-2R_{\hat0\hat n\hat0\hat m}\right)x^{\hat n}x^{\hat m}\textcolor[rgb]{0.00,0.59,0.00}{.}
\een

The curvature-tensor terms describe the lowest-order gravitational effects which a freely falling observer can measure locally.
For the special case of circular geodesics in the Schwarzschild field a Fermi normal tetrad has been obtained in \cite{PP}
\ben\lb{21a}
(e_{\hat0}^{\bar\alpha})=\left(\frac{\widetilde E}{X},0,0,\frac{\widetilde l}{r^2} \right),\qquad\\\lb{21b}
(e_{\hat1}^{\bar\alpha})=\left(-\frac{\widetilde l\sin(\alpha\varphi)}{r\sqrt{X}},\sqrt{X}\cos(\alpha\varphi),0,-\frac{\widetilde E\sin(\alpha\varphi)}{r\sqrt{X}} \right),\qquad\\\lb{21c}
(e_{\hat2}^{\bar\alpha})=\left(0,0,\frac{1}{r},0\right),\qquad\\\lb{21d}
(e_{\hat3}^{\bar\alpha})=\left(\frac{\widetilde l\cos(\alpha\varphi)}{r\sqrt{X}},\sqrt{X}\sin(\alpha\varphi),0,\frac{\widetilde E\cos(\alpha\varphi)}{r\sqrt{X}} \right).\qquad
\een
In these expressions we have introduced the abbreviations
\ben\lb{22}
\alpha=\sqrt{\frac{r-3\mathcal{M}}{r}},\qquad
X={1-\frac{2\mathcal{M}}{ r}}.
\een
At some initial time $(e_{\hat1}^{\bar\alpha})$ shows in radial direction and $(e_{\hat3}^{\bar\alpha})$ shows in
tangential direction, while $(e_{\hat2}^{\bar\alpha})$ is always perpendicular to the orbital plane.
One realizes that on the circular geodesic
\begin{equation}\label{geta}
g_{\alpha\beta}\frac{\partial x^{\alpha}}{\partial x^{\hat{\mu}}} \frac{\partial x^{\beta}}{\partial x^{\hat{\nu}}} = \eta_{\hat{\mu}\hat{\nu}}\,
\end{equation}
is valid, where $\eta_{\hat{\mu}\hat{\nu}}$ is the Minkowski metric.
The tetrads are parallel transported along the circular geodesic 
\begin{equation}\label{parallel}
\frac{D e^{\bar\mu}_{\hat{\alpha}}}{d \tau} = 0.
\end{equation}

The nonvanishing components of the Riemann tensor in Fermi normal coordinates are determined from
\ben\lb{20}
R_{\hat\mu\hat\nu\hat\sigma\hat\tau}=R_{\bar\alpha\bar\beta\bar\gamma\bar\delta}e_{\hat\mu}^{\bar\alpha}
e_{\hat\nu}^{\bar\beta}e_{\hat\sigma}^{\bar\gamma}e_{\hat\tau}^{\bar\delta}\,.
\een
Using
(\ref{18a}) -- (\ref{18c}) and (\ref{21a}) -- (\ref{20}) yields \cite{MM,PP,WZ,CK}
\ben\lb{23a}
&&R_{\hat0\hat1\hat0\hat1}=-\frac{\mathcal{M}[r+3(r-2\mathcal{M})\cos(2\alpha\varphi)]}{2(r-3\mathcal{M})r^3},
\\\lb{23b}
&&R_{\hat0\hat1\hat0\hat3}=-\frac{3\mathcal{M}(r-2\mathcal{M})\sin(2\alpha\varphi)}{2(r-3\mathcal{M})r^3},
\\\lb{23c}
&&R_{\hat0\hat2\hat0\hat2}=\frac{\mathcal{M}}{(r-3\mathcal{M})r^2},
\\\lb{23d}
&&R_{\hat0\hat3\hat0\hat3}=-\frac{\mathcal{M}[r-3(r-2\mathcal{M})\cos(2\alpha\varphi)]}{2(r-3\mathcal{M})r^3},
\\\lb{23e}
&&R_{\hat0\hat1\hat1\hat3}=\frac{3\mathcal{M}^\frac32\sqrt{r-2\mathcal{M}}\cos(\alpha\varphi)}{(r-3\mathcal{M})r^3},\\\lb{23f}
&&R_{\hat0\hat2\hat1\hat2}=-\frac{3\mathcal{M}^\frac32\sqrt{r-2\mathcal{M}}\sin(\alpha\varphi)}{(r-3\mathcal{M})r^3}.
\een
Furthermore, the following relationships hold \cite{CK}
\ben\lb{24a}
R_{\hat0\hat3\hat1\hat3}=-R_{\hat0\hat2\hat1\hat2},\qquad R_{\hat0\hat2\hat2\hat3}=-R_{\hat0\hat1\hat1\hat3},
\\\lb{24b}
R_{\hat1\hat2\hat1\hat2}=-R_{\hat0\hat3\hat0\hat3}, \qquad R_{\hat1\hat3\hat1\hat3}=-R_{\hat0\hat2\hat0\hat2},
\\\lb{24c}
R_{\hat1\hat2\hat2\hat3}=-R_{\hat0\hat1\hat0\hat3}, \qquad R_{\hat2\hat3\hat2\hat3}=-R_{\hat0\hat1\hat0\hat1}.
\een

Via $\varphi$ the components of $R_{\hat\mu\hat\nu\hat\sigma\hat\tau}$ depend on $\tau$.
Moreover, because of the nonvanishing component $R_{\hat{0}\hat{1}\hat{0}\hat{3}}$ the second-order contribution to $g_{\hat0\hat0}$ is not diagonal.
A simpler structure can be obtained by performing
 a tetrad rotation around the $x^{\hat2}$ direction according to
 \begin{eqnarray}\label{e1}
\textbf{\underline{E}}_{\bar{1}} &=&
\mathbf{\underline{e}}_{\hat{1}} \cos\alpha\varphi
+ \mathbf{\underline{e}}_{\hat{3}} \sin\alpha\varphi \nonumber \\
\textbf{\underline{E}}_{\bar{3}} &=&
- \mathbf{\underline{e}}_{\hat{1}} \sin\alpha\varphi
+ \mathbf{\underline{e}}_{\hat{3}} \cos\alpha\varphi\,.
\end{eqnarray}
Now $(E_{\hat1}^{\bar\alpha})$ always shows in radial direction and $(E_{\hat3}^{\bar\alpha})$ always shows in
tangential direction.
Explicitly
\begin{eqnarray}
% \nonumber to remove numbering (before each equation)
E_{\hat{0}}^{\bar0}&=& \sqrt{\frac{r}{r-3\mathcal{M}}}\,,\quad E_{\hat{0}}^{\bar3}= \frac{1}{r}\sqrt{\frac{\mathcal{M}}{r-3\mathcal{M}}}\,,\\
E_{\hat{1}}^{\bar1} &=& \left(1 - \frac{2\mathcal{M}}{r}\right)^{1/2}\,,\quad  E_{\hat{2}}^{\bar2}  = \frac{1}{r}\,,\\
 E_{\hat{3}}^{\bar0}  &=& \sqrt{\frac{\mathcal{M}r}{\left(r-2\mathcal{M}\right)\left(r-3\mathcal{M}\right)}},\quad  E_{\hat{3}}^{\bar3} = \frac{1}{r}\sqrt{\frac{r-2\mathcal{M}}{r-3\mathcal{M}}}\,.\nonumber\\
\end{eqnarray}
The tetrads $\mathbf{\underline{e}}_{\hat{0}}$ and $\mathbf{\underline{e}}_{\hat{2}}$ are unchanged.
In this frame the curvature-tensor components take a simpler form:
\ben\lb{26a}
&&R_{0101}=-\frac{\mathcal{M}}{r^3}\frac{2r-3\mathcal{M}}{r-3\mathcal{M}}=-R_{2323},
\\\lb{26b}
&&R_{0303}=\frac{\mathcal{M}}{r^3}=-R_{1212},
\\\lb{26c}
&&R_{0113}=\frac{3\mathcal{M}}{r^3}\frac{\sqrt{\mathcal{M}(r-2\mathcal{M})}}{r-3\mathcal{M}}=
R_{0232},
\\\lb{26d}
&&R_{0202}=\frac{\mathcal{M}}{r^3}\frac{r}{r-3\mathcal{M}}=-R_{1313}.
\een

The price to pay for this simplification is that the new tetrads are no longer parallel transported.
Combining $\frac{D\mathbf{\underline{e}}_{\hat{1}}}{d\tau} =0$ and
$\frac{D\mathbf{\underline{e}}_{\hat{3}}}{d\tau} =0$  from (\ref{parallel}) with (\ref{e1})
provides us with
\begin{equation}\label{e2}
\frac{D\mathbf{\underline{E}}_{\bar{1}}}{d\tau} = \sqrt{\frac{\mathcal{M}}{r^{3}}}\mathbf{\underline{E}}_{\bar{3}}\,,\qquad
\frac{D\mathbf{\underline{E}}_{\bar{3}}}{d\tau} = -\sqrt{\frac{\mathcal{M}}{r^{3}}}\mathbf{\underline{E}}_{\bar{1}}\,.
\end{equation}
The transformation describes a rotation
with frequency $\sqrt{\frac{\mathcal{M}}{r^{3}}}$. As a result, there appear nonvanishing Christoffel symbols
on the geodesic together with mixed spacetime terms linear in $x^i$ in the metric.
This corresponds to what is called a proper reference frame in Ref. \cite{MTW} (see also Chap. 21.2 in Ref. \cite{St}) in the absence of acceleration. But while the proper reference system in \cite{MTW,St} is restricted to linear deviations from the central geodesic, our analysis includes the deviations up to second order.
The relevant relations are (cf. Refs. \cite{MTW,St})
\begin{equation}\label{g1}
\Gamma^{a}_{0n} = \varepsilon_{n\ m}^{\ \,a}\omega^{m}\,,\qquad g_{0b,n} = - \varepsilon_{bnm}\omega^{m}\,.
\end{equation}
In our case
\begin{equation}\label{Gamma}
\Gamma^{1}_{03 } = \varepsilon_{3\ 2}^{\ \,1}\,\omega =  \omega\,,\qquad
\Gamma^{3}_{01 } = -\varepsilon_{1\ 2}^{\ \,3}\,\omega = -\omega\,.
\end{equation}
It follows that
\begin{equation}\label{g2}
g_{01,3} = - \varepsilon_{132}\omega = \omega\,,\qquad g_{03,1} = - \varepsilon_{312}\omega^{m}
= -\omega\,.
\end{equation}
With $\omega = \sqrt{\frac{\mathcal{M}}{r^{3}}}$ we have
\begin{equation}\label{g3}
g_{01} = \sqrt{\frac{\mathcal{M}}{r^{3}}}x^{3}\,,\qquad g_{03} = -\sqrt{\frac{\mathcal{M}}{r^{3}}}x^{1}\,.
\end{equation}

Once we know the curvature-tensor components in this new frame, we can obtain the metric tensor up to second order in the deviations. The focus in the present paper is on the component $g_{00}$ since it is this component which appears in Tolman's and Klein's laws. We find
\ben\no
g_{00}&=&-1+\frac{\mathcal{M}}{r^3}\bigg[\frac{2r-3\mathcal{M}}{r-3\mathcal{M}}
\left(x^1\right)^2
\\\lb{27a}&&\qquad \qquad\quad - \frac{r}{r-3\mathcal{M}}\left(x^2\right)^2-\left(x^3\right)^2\bigg].
\een

\section{The temperature profile}
\label{profile}
Now we combine the metric structure (\ref{27a}) with Tolman's relation (\ref{Tolman}). As a result we obtain a  parabolic temperature profile in the vicinity of the central geodesic 
\ben\no
T=\frac{T_0}{\sqrt{-g_{00}}}\approx T_0\bigg\{1+\frac{\mathcal{M}}{2r^3}\bigg[\frac{2r-3\mathcal{M}}{r-3\mathcal{M}}
\left(x^1\right)^2
\\\lb{30}
-\frac{r}{r-3\mathcal{M}}\left(x^2\right)^2-\left(x^3\right)^2\bigg]\bigg\}\,,
\een
where $T_{0}$ is the constant equilibrium temperature on the geodesic.
Obviously, the temperature variations are different in different directions off the geodesic.
The parabolic structure of the temperature distribution implies that the spatial gradient of the temperature is linear in the deviation from the central geodesic. Consequently, the temperature gradient vanishes on the geodesic itself. Therefore, on this  geodesic, and only there, the second
of the equilibrium conditions (\ref{10a}) consistently reduces to the equation $\dot{U}^{\mu}=0$ for geodesic fluid motion. Off the central geodesic the equilibrium fluid motion is nongeodesic.
However, since the deviation is linear in the distance the situation simplifies as we shall discuss in the following section.

\section{Geodesic deviation}
\label{geodesic}

Apparently, the second equilibrium condition (\ref{10a}) is no longer compatible with a geodesic motion  $\dot{U}^{\mu}=0$ in the vicinity of the circular geodesic. But since the terms that ``perturb" the geodesic behavior are linear in the distance, the equation of geodesic deviation turns out to be applicable for our problem. The reason is that these perturbing terms lead to higher-order corrections for the geodesic deviation. In the following we demonstrate this in some detail.

Quite generally, for a vector $\xi^\alpha$ orthogonal to the geodesic the equation for the geodesic deviation is
\ben\lb{28a}
\frac{D^2\xi^\alpha}{d\tau^2}+{R^\alpha}_{\gamma\mu\nu}U^\gamma U^\nu \xi^\mu=0,
\een
where the explicit form of the first term of this equation reads
\ben\no
\frac{D^2\xi^\alpha}{d\tau^2}=\frac{d^2\xi^\alpha}{d\tau^2}+\Gamma^\alpha_{\beta\gamma,\rho}\,
\xi^\beta\frac{dx^\rho}{d\tau}\frac{dx^\gamma}{d\tau}
+2\Gamma^\alpha_{\beta\gamma}\frac{d\xi^\beta}{d\tau}\frac{dx^\gamma}{d\tau}
\\\lb{28b}
+\underline{\Gamma^\alpha_{\beta\gamma}\xi^\beta\frac{d^2x^\gamma}{d\tau^2}}
+\Gamma^\alpha_{\beta\gamma}\Gamma^\beta_{\rho\sigma}\xi^\rho\frac{dx^\sigma}{d\tau}
\frac{dx^\gamma}{d\tau}.\qquad
\een

Let us analyze the underlined term in (\ref{28b}).
Writing the second equilibrium condition (\ref{10a}) as
\ben\lb{29}
\dot U^\gamma = \frac{DU^\gamma}{d\tau} = \frac{D^2x^\gamma}{d\tau^2}=- \frac{c^{2}}{T}\nabla^{\gamma}T\,,
\een
and
\begin{equation}\label{x1}
\frac{d^2x^\gamma}{d\tau^2} = - \Gamma^{\gamma}_{\kappa\lambda}U^{\kappa}U^{\lambda} - \frac{c^{2}}{T}\nabla^{\gamma}T\,,
\end{equation}
the underlined term is
\begin{equation}\label{x2}
\Gamma^\alpha_{\beta\gamma}\xi^\beta\frac{d^2x^\gamma}{d\tau^2} = - \Gamma^\alpha_{\beta\gamma}\xi^\beta
\left[\Gamma^{\gamma}_{\kappa\lambda}U^{\kappa}U^{\lambda} + \frac{c^{2}}{T}\nabla^{\gamma}T\right]\,.
\end{equation}
Since $\nabla_{\gamma}T$ is linear in the distance and the entire term is already linear in $\xi^{\beta}$, the temperature gradient gives rise to a second-order contribution.
Hence, up to linear order we get the relationship
\ben\lb{32}
\Gamma_{\beta\gamma}^\alpha\xi^\beta\frac{d^2x^\gamma}{d\tau^2}=-
\Gamma_{\beta\gamma}^\alpha\xi^\beta\Gamma^\gamma_{\kappa\lambda}U^\kappa U^\lambda.
\een
Furthermore, in the terms that multiply
$\xi^{\alpha}$ we may approximate (\ref{11a}) by
$(U^\mu)=\left(c,\bf0\right)$
since in our metric $g_{00} = -1 + \mathcal{O}(x^{2})$ and any correction to $g_{00} = -1$ would lead to higher-order terms.
In this case (\ref{28b}) together with (\ref{32}) yields
\ben\no
\frac{D^2\xi^\alpha}{d\tau^2}=\frac{d^2\xi^\alpha}{d\tau^2}+c^2\Gamma^\alpha_{\beta0,0}\,
\xi^\beta+2c\Gamma^\alpha_{\beta0}\frac{d\xi^\beta}{d\tau}
\\\lb{33a}
-c^2\Gamma_{\beta\gamma}^\alpha\xi^\beta\Gamma^\gamma_{00}
+c^2\Gamma^\alpha_{\beta0}\Gamma^\beta_{\rho0}\xi^\rho,
\een
while the second term in (\ref{28a}) reduces to
\ben\lb{33b}
{R^\alpha}_{\gamma\mu\nu}U^\gamma U^\nu \xi^\mu=c^2g^{\alpha\beta}R_{\beta0\mu0}\xi^\mu.
\een

Now we identify the spatial components of $\xi^a$ with the components $x^a$ of our tetrad system.
Then, from  (\ref{28a}) together with (\ref{33a}) and (\ref{33b}), we obtain the following form of the equation for the geodesic deviation:
\ben\no
\frac{d^2x^a}{d\tau^2}+2c\Gamma^a_{\beta0}\frac{dx^\beta}{d\tau}
-c^2\Gamma_{\beta\gamma}^ax^\beta\Gamma^\gamma_{00}
+c^2\Gamma^a_{\beta0}\Gamma^\beta_{\rho0}x^\rho
\\\lb{33c}
+c^2g^{a\beta}R_{\beta0\mu0}x^\mu=0.
\een
Note that the time derivative of the Christoffel symbols vanishes, since the metric is static. The temporal part of the equation for the geodesic deviation is identically satisfied.

 Consequently, by taking into account the expressions (\ref{26a}) - (\ref{26d}) for the components of the Riemann tensor and those for the Christoffel symbols in (\ref{Gamma}), we arrive at the linear-order system of equations
\ben\lb{34b}
\frac{d^2x^1}{d\tau^2}-2c\sqrt{\frac{\mathcal{M}}{r^3}}\frac{dx^3}{d\tau}
-3c^2\frac{\mathcal{M}}{r^3}\frac{r-2\mathcal{M}}{r-3\mathcal{M}}x^1=0,\\\lb{34c}
\frac{d^2x^2}{d\tau^2}
+c^2\frac{\mathcal{M}}{r^3}\frac{r}{r-3\mathcal{M}}x^2=0,\\\lb{34d}
\frac{d^2x^3}{d\tau^2}+2c\sqrt{\frac{\mathcal{M}}{r^3}}\frac{dx^1}{d\tau}=0.
\een

The system (\ref{34b}) - (\ref{34d}) coincides with the system derived for the motion of a test particle in Ref. \cite{Sh}.

Equation (\ref{34c}) decouples from the other equations and describes oscillations in the component  $x^2$. The real solution can be written as
\be
x^2=x_0^2\,\sin\left(\Omega\,\tau\right),\quad \hbox{with}\quad \Omega=\sqrt{\frac{GM}{r^3}\frac{r}{r-3\mathcal{M}}},
\ee{35a}
where $\Omega$ denotes the frequency of the harmonic motion of the $x^2$ component.
This frequency coincides with the orbital frequency $\omega_{\varphi}$ in (\ref{ophi}).
But in the present context it characterizes an oscillation perpendicular to the orbital plane.

The coupled system of equations (\ref{34b}) and (\ref{34d}) has oscillatory solutions as well which can be written as
\ben\lb{35b}
x^1&=&x_0^1\,\sin\left(\omega\,\tau\right),\\ \lb{35c} x^3&=&2\sqrt{\frac{r-3\mathcal{M}}{r-6\mathcal{M}}}x_0^1\,\cos\left(\omega\,\tau\right).
\een
Here the oscillation frequency is
\ben\lb{35d}
\omega=\sqrt{\frac{GM}{r^3}\frac{r-6\mathcal{M}}{r-3\mathcal{M}}}.
\een
This frequency coincides with the radial frequency $\omega_{r}$ in (\ref{or}). But now it characterizes also oscillations in tangential direction.
Hence, in the $(x^1,x^3)$ plane the motion is described by an ellipse.
Different from the situation in terms of Schwarzschild coordinates in Sec.~\ref{Schwarzschild}, there is no precession in the orbital plane for a comoving observer.

As already discussed in Sec.~\ref{Schwarzschild}, the frequencies $\Omega$ and $\omega$ only coincide in the limit $\mathcal{M} \ll r$ where
\begin{equation}\label{w1}
\omega_{N} = \Omega_{N} = \sqrt{\frac{GM}{r^{3}}}.
\end{equation}
Their ratio is given by
\ben\lb{36}
\frac\omega\Omega=\sqrt{1-\frac{6\mathcal{M}}{r}}.
\een
Oscillations with frequencies $\Omega$ and $\omega$ have been derived for geodesic particle motion in the Schwarzschild field by Shirokov \cite{Sh}. The new feature here is that, via Tolman's relation, these frequencies are also relevant for the gas temperature and other thermodynamic quantities as we shall discuss in the following section.

As pointed out in Ref. \cite{Sh}, the first-order corrections to the Newtonian frequency are different for $\Omega$ and $\omega$:
\begin{equation}\label{w2}
\omega \approx \omega_{N}\left(1 - \frac{3}{2}\frac{\mathcal{M}}{r}\right)\,,\quad
\Omega \approx \Omega_{N}\left(1 + \frac{3}{2}\frac{\mathcal{M}}{r}\right)\,.
\end{equation}

 On the other hand, stable circular orbits only exist for $r> 6\mathcal{M}$ \cite{Wald}. In the limit $r \rightarrow  6\mathcal{M}$ one has $\omega\rightarrow 0$ and $\Omega\rightarrow \sqrt{2}\Omega_{N}$.
The oscillations are frozen in the $x^{1}-x^{3}$ plane, while they continue with the frequency $\sqrt{2}\Omega_{N}$ in the $x^{2}$ plane.
In the interval $3\mathcal{M} < r < 6\mathcal{M}$ there exist unstable circular trajectories. Since $\omega$ becomes imaginary in this region, there are exponential instabilities in the $x^{1}-x^{3}$ plane. The oscillation frequency in the $x^{2}$ direction increases to very large values as the limit $r = 3\mathcal{M}$ is approached.

\section{Thermodynamic properties in circular geodesic motion}
\label{thermodynamic}
In this section we shall analyze the thermodynamic fields of a rarefied gas  as measured by an observer in circular geodesic motion.

\subsection{Temperature oscillations}

From (\ref{30}) with (\ref{35a}) -- (\ref{35d}) we obtain the temperature profile
\ben\no
\frac{T-T_0}{T_0}=\Delta(\tau)=\frac{\mathcal{M}}{2r^3}\bigg\{\left(x_0^1\right)^2\bigg[A\sin^2\left(\omega\tau\right)
\\\lb{37a}
-B\cos^2\left(\omega\tau\right)\bigg]
-C\sin^2\left(\Omega\tau\right)\bigg\},
\een
where we have introduced the abbreviations
\be
A\equiv \frac{2r-3\mathcal{M}}
{r-3\mathcal{M}},\quad B\equiv 4\frac{r-3\mathcal{M}}
{r-6\mathcal{M}}, \quad C\equiv \frac{r}{r-3\mathcal{M}}\,.
\ee{a}

Since the solutions of the system (\ref{34b}) -- (\ref{34d}) enter quadratically, the oscillation frequencies are doubled.  The explicit expression is
\begin{eqnarray}
\label{deltatau}
\Delta(\tau)&=&\frac{\mathcal{M}}{4r^3}\left(x_0^1\right)^2\bigg\{\left(A-B\right)
 -\left(A+B\right)\cos{\left(2\omega\tau\right)}  \nonumber\\\lb{dt}
&&- C\left(\frac{x_0^2}{x_0^1}\right)^2  + C\left(\frac{x_0^2}{x_0^1}\right)^2\cos{\left(2\Omega\tau\right)}\bigg\}.
\end{eqnarray}
The temperature field oscillates in the $(x^1,x^3)$ plane with the frequency $2\omega$ and different amplitudes, while in the $x^2$ direction the oscillation frequency is $2\Omega$. In the Newtonian limit we have
\begin{eqnarray}
\Delta_{N}(\tau)&=&-\frac{\mathcal{M}}{4r^3}\left(x_0^1\right)^2\left\{2
 +\left(\frac{x_0^2}{x_0^1}\right)^2  \right.\nonumber\\\lb{dn}
&&\left. +  \left[6 - \left(\frac{x_0^2}{x_0^1}\right)^2\right]\cos{\left(2\sqrt{\frac{GM}{r^3}}\tau\right)} \right\}.
\end{eqnarray}
The lowest-order general-relativistic corrections modify the amplitudes according to
\begin{equation}\label{ap1}
A-B\approx - 2 - 9\frac{\mathcal{M}}{r},\quad A+B \approx 6 + 15\frac{\mathcal{M}}{r},\quad C\approx 1+ 3\frac{\mathcal{M}}{r}.
\end{equation}
The more interesting effect is a modulation of the frequencies
\begin{equation}\label{b}
\cos{\left(2\omega\tau\right)} \approx \cos{\left(2\omega_{N}\left(1-\frac{3}{2}\frac{\mathcal{M}}{r}\right)\tau\right)},
\end{equation}
\begin{equation}\label{c}
\cos{\left(2\Omega\tau\right)} \approx \cos{\left(2\Omega_{N}\left(1+\frac{3}{2}\frac{\mathcal{M}}{r}\right)\tau\right)}.
\end{equation}
It follows that the oscillation periods are different in different directions. Equivalently,
this modulation can be expressed as
\begin{eqnarray}
\cos{\left(2\omega\tau\right)} &\approx & \cos{\left(2\omega_{N}\tau\right)}\cos{\left(3\frac{\mathcal{M}}{r}\omega_{N}\tau\right)}\nonumber  \\
  && + \sin{\left(2\omega_{N}\tau\right)}\sin{\left(3\frac{\mathcal{M}}{r}\omega_{N}\tau\right)}
\end{eqnarray}
and
\begin{eqnarray}
\cos{\left(2\Omega\tau\right)} &\approx & \cos{\left(2\Omega_{N}\tau\right)}\cos{\left(3\frac{\mathcal{M}}{r}\Omega_{N}\tau\right)}\nonumber  \\
  && -\sin{\left(2\Omega_{N}\tau\right)}\sin{\left(3\frac{\mathcal{M}}{r}\Omega_{N}\tau\right)}.
\end{eqnarray}
The Newtonian oscillations with frequency $2\omega_{N} = 2\Omega_{N}$ are modulated by the very small frequency
$3\frac{\mathcal{M}}{r}\omega_{N}$.

At lowest order, the frequency difference $\Delta\omega$ is
\begin{equation}\label{zz}
\Delta\omega = 2\left(\Omega -\omega\right) = 6 \frac{\left(GM\right)^{3/2}}{c^{2}r^{5/2}}.
\end{equation}

The deviations from Newtonian behavior are more drastic for strong fields. Let us consider the case $r=7\mathcal{M}$ which is well in the range of stable circular orbits. In this case
\begin{equation}\label{ap2}
A = \frac{11}{4},\quad B = 16, \quad C = \frac{7}{4},
\end{equation}
 and the frequencies $\omega$ and $\Omega$ become
\begin{equation}\label{w3}
\omega = \frac{\omega_{N}}{2}\quad \mathrm{and}\quad \Omega = \frac{\sqrt{7}}{2}\omega_{N},
\end{equation}
respectively. The frequency $\omega$ is considerably smaller than the Newtonian frequency, while $\Omega$ is considerably larger. The temperature field oscillates with $2\omega = \omega_{N}$ and $2\Omega = \sqrt{7}\Omega_{N}$. The oscillation periods differ by a factor of $\sqrt{7}$.
The frequency difference is
\begin{equation}\label{zz2}
\Delta\omega = \left(\sqrt{7} -1\right)\omega_{N} \approx 1.65\omega_{N}\,.
\end{equation}
Likewise we obtain the values for $r=10\mathcal{M}$:
\ben
A = \frac{17}{7},\quad B = 7, \quad C = \frac{10}{7},\\
\omega = \frac{2}{\sqrt{7}}\omega_{N}, \qquad\Omega = \sqrt\frac{10}{7}\omega_{N}.
\een
Here we find
\begin{equation}\label{zz1}
\Delta\omega \approx 0.88\omega_{N}\,.
\end{equation}
Interestingly, the ratio between the frequencies for $r=10\mathcal{M}$ is $\Omega/\omega = \sqrt{2.5} \approx 1.58$
which is close to the frequently observed twin-peak ratio 3/2 in the power spectra of x-ray binaries \cite{Abramowicz}.

As already mentioned, when $r$ approaches $6\mathcal{M}$ the frequency $\omega$ approaches $0$ while the coefficient $B$ diverges. This indicates the onset of an instability. In the range $3\mathcal{M} < r < 6\mathcal{M}$ the coefficient $B$ is negative and we have an imaginary $\omega$, indicating an exponential instability. As $r$ approaches $3\mathcal{M}$, the oscillation frequency in the $x^{2}$ direction becomes infinitely large.

\subsection{Thermodynamic functions at equilibrium}
From Klein's law ($\mu\sqrt{-g_{00}}$= constant) it follows that the chemical potential has the same oscillatory character as the temperature field, namely,
\ben\lb{38}
\frac{\mu-\mu_0}{\mu_0}=\Delta(\tau)\textcolor[rgb]{0.00,0.59,0.00}{.}
\een

The energy per particle [left-hand expression of (\ref{5a})] is only a function of the temperature so that we can write
\ben\lb{39a}
\frac{\e-\e_0}{kT_0}=\frac{\c_v^0}{k}\Delta(\tau),
\een
where $\c_v$ is the heat capacity per particle at constant volume 
\ben\no
\c_v^0=\frac{\partial \e}{\partial T}\Bigg\vert_{T_0}=k\z_0^2\left[1+\frac{5}{\z_0}\frac{K_3(\z_0)}{K_2(\z_0)}\right.
\\\lb{39b}\left.
-\left(\frac{K_3(\z_0)}{K_2(\z_0)}\right)^2-\frac{1}{\z_0^2}\right].
\een

According to (\ref{6a}) and the left-hand expression of (\ref{9}) the particle number density $\n$ is only a function of the temperature as well 
\be
\n={4\pi}m^2c\frac{K_2(\z)}{\z}e^{\frac{\mu}{kT}-1}={4\pi}m^2c\frac{K_2(\z)}{\z}e^{\frac{\mu_0}{kT_0}-1},
\ee{40a}
and  we may express it as
\be
\n=\n_0+\frac{\partial \n}{\partial T}\Bigg\vert_{T_0}(T-T_0).
\ee{40b}
From this equation and from (\ref{37a}) and (\ref{38}) we obtain the behavior of the particle number density in the  vicinity of the circular geodesic:
\be
\frac{\n-\n_0}{\n_0}=\left[\z_0\frac{K_3(\z_0)}{K_2(\z_0)}-1\right]\Delta(\tau).
\ee{40c}

The oscillations of the pressure field  follow from the equation of state $\p=\n kT$, (\ref{37a}), (\ref{40c}) and read
\be
\frac{\p-\p_0}{\p_0}=\z_0\frac{K_3(\z_0)}{K_2(\z_0)}\Delta(\tau).
\ee{40d}

With  (\ref{5b}) and (\ref{6a}) the entropy per particle is
\ben\lb{41}
\s=k\left[\z\frac{K_3(\z)}{K_2(\z)}-\frac{\mu_0}{kT_0}\right].
\een
Following similar steps as those which led us to (\ref{40b}), we find
\be
\frac{\s-\s_0}{\s_0}=-\left[\frac{\frac{K_3(\z_0)}{K_2(\z_0)}
\left(\z_0\frac{K_3(\z_0)}{K_2(\z_0)}-4\right)-\z_0}{\frac{K_3(\z_0)}{K_2(\z_0)}
-\frac{\mu_0}{mc^2}}\right]\Delta(\tau).
\ee{42}

As an example we determine the  equilibrium fields in the nonrelativistic  limit $\z\gg1$. Under this condition the fields (\ref{5a}) -- (\ref{6a}) and (\ref{39b})   up to the order $1/\z$ read
 \ben\lb{43a}
 \e_0=mc^2+\frac32 kT_0\left[1+\frac{5}{4\z_0}\right],\quad\\\lb{43b}
 \s_0=k\left[\ln\frac{T_0^\frac32}{\n_0}+\frac32\ln(2\pi  ek m)+\frac{15}{4\z_0}\right],\quad\\\lb{43c}
 \mu_0=mc^2+kT_0\left[\ln\frac{e\n_0}{T_0^\frac32}-\frac32\ln(2\pi k m)-\frac{15}{8\z_0}\right],\quad\\\lb{43d}
 \c_v^0=\frac32 k\left[1+\frac{5}{2\z_0}\right].\quad
 \een

With (\ref{43a}) -- (\ref{43d}) the relations (\ref{39a}), (\ref{40d}) and (\ref{42}) reduce to
 \ben\lb{44a}
 \frac{\e-\e_0}{kT_0}&=&\frac32 \left[1+\frac{5}{2\z_0}\right]\Delta(\tau),\\\lb{44b}
 \frac{\p-\p_0}{\p_0}&=&\left[1+\frac{5}{2\z_0}\right]\z_0\Delta(\tau)
  \een
  and
  \begin{equation}\label{44c}
  \frac{\s-\s_0}{\s_0}=\frac1{\ln\frac{e\n_0}{T_0^\frac32}
 -\frac32\ln(2\pi  k m)}\z_0\Delta(\tau),
  \end{equation}
  respectively.

 Hence, apart from the factors, the energy per particle shows the same dependence on $\tau$ as the temperature and the chemical potential.
 However, the oscillation amplitudes of the pressure and the entropy per particle are larger, since they are multiplied by  $\z_0=mc^2/kT_0$ which has a big value in this limit.

\subsection{Analysis of temperature oscillations}

Now let us analyze the temperature oscillations of a gas, e.g. in a spacecraft, at low altitudes in orbits around the planets of the Solar System, where we can approximate the orbit radius $r$ by the radius $R$ of the massive object, i.e., $r\approx R$. In the first two columns of Table \ref{tab1} we  specify the radii and the masses of the planets.  The ratios $\mathcal{M}/R=GM/Rc^2$ in the third column are sufficiently small so that we can use the approximation (\ref{dn}) for the oscillation $\Delta(\tau)$.
From the fourth column we infer that the  frequencies $\omega_N=\sqrt{{GM}/{R^3}}$ for the four outer planets are about one-half of the ones for the four inner planets.
The fifth  column contains the corresponding oscillation amplitudes ${GM} /{R^3c^2}$.
\ \\

%%%%%%%%%%%%%%%%%%%%%%%%%%%%%%%%%%%%%%%%%%%%%%%%%%%%%%%%%%%%%%%%%%%%%%%%%%%%%%%%%%%%%%%%%%%%%%%%%%%%%%%%%%%%%%%%%%%%%%%%%
%%%%%%%%%%%%%%%%%%%%%%%%%%%%%%%%%%%%%%%%%%%%%%%%%%%%%%%%%%%%%%%%%%%%%%%%%%%%%%%%%%%%%%%%%%%%%%%%%%%%%%%%%%%%%%%%%%%%%%%
%\begin{widetext}
\begin{table}[ht]
\begin{tiny}
\caption{Parameters for planets in the Solar System}
\begin{tabular}{ccccccc}
\hline
 &$R$   &$M$ &$GM/Rc^2$  &$\omega_N$  &${GM}/ {R^3c^2}$  \\
&{\rm m}&{\rm kg}& &{\rm s$^{-1}$}&{\rm m$^{-2}$} \\
\hline
Mercury &$2.44\times 10^6$ & $3.30\times
10^{23}$ &$1.00\times10^{-10}$&  $1.23\times 10^{-3}$   & $1.69\times10^{-23}$\\
Venus &$6.05\times 10^6$ & $4.87\times
10^{24}$ &$5.97\times10^{-10}$&  $1.21\times 10^{-3}$   & $1.63\times10^{-23}$\\
Earth &$6.38\times 10^6$ & $5.97\times
10^{24}$ &$6.95\times10^{-10}$&  $1.24\times 10^{-3}$   & $1.71\times10^{-23}$\\
Mars & $3.39\times10^6$ & $6.41\times10^{23}$ &$1.41\times10^{-10}$& $1.05\times10^{-3}$    & $1.22\times10^{-23}$\\
Jupiter & $7.00\times 10^7$ & $1.90\times10^{27}$ &$2.01\times10^{-8}$& $6.09\times10^{-4}$ & $4.13\times10^{-24}$\\
Saturn & $5.82\times 10^7$ & $5.68\times10^{26}$ &$7.25\times10^{-9}$& $4.38\times 10^{-4}$ & $2.14\times10^{-24}$\\
Uranus &$2.54\times 10^7$ & $8.68\times
10^{25}$ &$2.54\times 10^{-9}$&  $5.96\times10^{-4}$   & $3.95\times10^{-24}$\\
Neptune &$2.46\times 10^7$ & $1.02\times
10^{26}$ &$3.09\times10^{-9}$&  $6.77\times 10^{-4}$   & $5.09\times10^{-24}$\\
\hline
\end{tabular}
\label{tab1}
\end{tiny}
\end{table}
%\end{widetext}
%%%%%%%%%%%%%%%%%%%%%%%%%%%%%%%%%%%%%%%%%%%%%%%%%%%%%%%%%%%%%%%%%%%%%%%%%%%%%%%%%%%%%%%%%%%%%%%%%%%%%%%%%%%%%%%%%%%%%%
%%%%%%%%%%%%%%%%%%%%%%%%%%%%%%%%%%%%%%%%%%%%%%%%%%%%%%%%%%%%%%%%%%%%%%%%%%%%%%%%%%%%%%%%%%%%%%%%%%%%%%%%%%%%%%%%%%%

In Fig. \ref{fig1} we show the temperature oscillations $\Delta_N(\tau)$ of a gas in circular motion around Earth, Mars and Saturn as functions of the proper time $\tau$. As an example we have taken $x_0^1=x_0^2=1$. The curves complement the data of the table, indicating that the frequencies for Earth and Mars are practically the same.  The amplitudes are different, however, the amplitude of the latter being smaller than that of the former. For Saturn both the oscillation frequency and the amplitude are smaller than the corresponding quantities for Earth and for Mars.
Note that the fractional amplitudes of the oscillations are of the order of $10^{-23}$; i.e., they are very small.\\
\ \\
\ \\
%%%%%%%%%%%%%%%%%%%%%%%%%%%%%%%%%%%%%%%%%%%%%%%%%%%%%%%%%%%%%%%%%%%%%%%%%%%%%%%%%%%%%%%%%%%%%%%%%%%%%%%%%%%%%%%%%%%%
%%%%%%%%%%%%%%%%%%%%%%%%%%%%%%%%%%%%%%%%%%%%%%%%%%%%%%%%%%%%%%%%%%%%%%%%%%%%%%%%%%%%%%%%%%%%%%%%%%%%%%%%%%%%%%%%%%%
\begin{figure}[ht]
 \includegraphics[width=0.5\textwidth]{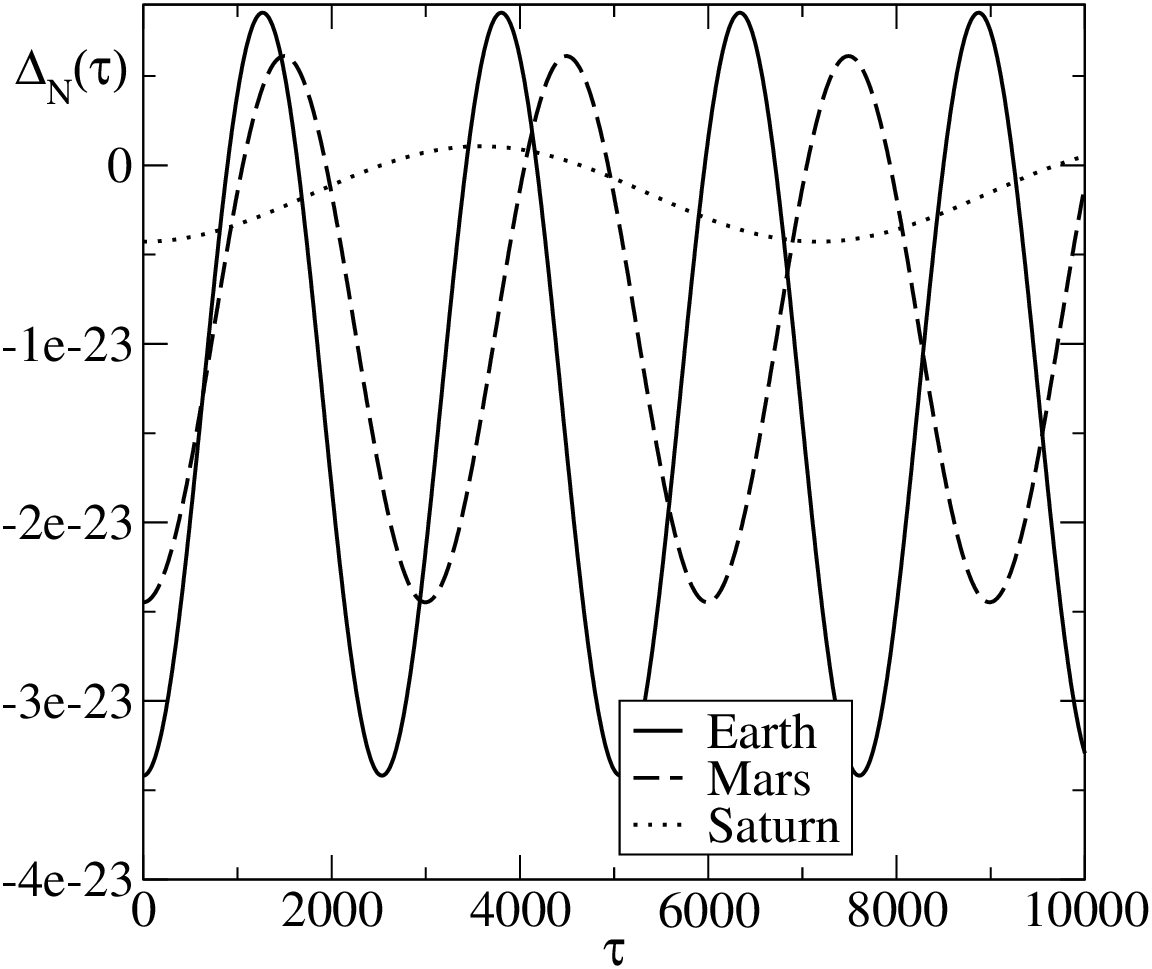}
\caption{Temperature oscillations of a gas  in circular motion around Earth, Mars and Saturn.}
\lb{fig1}
\end{figure}
%%%%%%%%%%%%%%%%%%%%%%%%%%%%%%%%%%%%%%%%%%%%%%%%%%%%%%%%%%%%%%%%%%%%%%%%%%%%%%%%%%%%%%%%%%%%%%%%%%%%%%%%%%%%%%%%%%%%%%%%
%%%%%%%%%%%%%%%%%%%%%%%%%%%%%%%%%%%%%%%%%%%%%%%%%%%%%%%%%%%%%%%%%%%%%%%%%%%%%%%%%%%%%%%%%%%%%%%%%%%%%%%%%%%%%%%%%%%%%%

As a second application we  analyze the previously considered case $r=7\mathcal{M}$ which corresponds to strong fields. If the massive object has a radius of the order of Earth's radius, the Newtonian frequency is about $w_N\approx 17.77$ Hz. The temperature oscillations $\Delta(\tau)$ for this case are plotted in Fig. \ref{fig2} as function of the proper time $\tau$. As to be expected,  both the oscillation frequencies and the amplitudes here are larger than those for the planets of the Solar System. Moreover, the frequency difference becomes clearly visible.

The oscillation amplitudes for the pressure and the entropy per particle are larger than those for the temperature, the energy per particle and the chemical potential, since the amplitudes of the former are multiplied by $\z_0=mc^2/kT$.
For a hydrogen gas H$_2$ at a temperature of 300 K this factor is about $\z_0\approx7.2\times 10^{10}$.
Hence the oscillations of the pressure and the entropy per particle are more pronounced than those for the temperature, the energy per particle and the chemical potential.\\
\ \\
%%%%%%%%%%%%%%%%%%%%%%%%%%%%%%%%%%%%%%%%%%%%%%%%%%%%%%%%%%%%%%%%%%%%%%%%%%%%%%%%%%%%%%%%%%%%%%%%%%%%%%%%%%%%%%%%%%
%%%%%%%%%%%%%%%%%%%%%%%%%%%%%%%%%%%%%%%%%%%%%%%%%%%%%%%%%%%%%%%%%%%%%%%%%%%%%%%%%%%%%%%%%%%%%%%%%%%%%%%%%%%%%%%%
\ \\
\begin{figure}[ht]
 \includegraphics[width=0.5\textwidth]{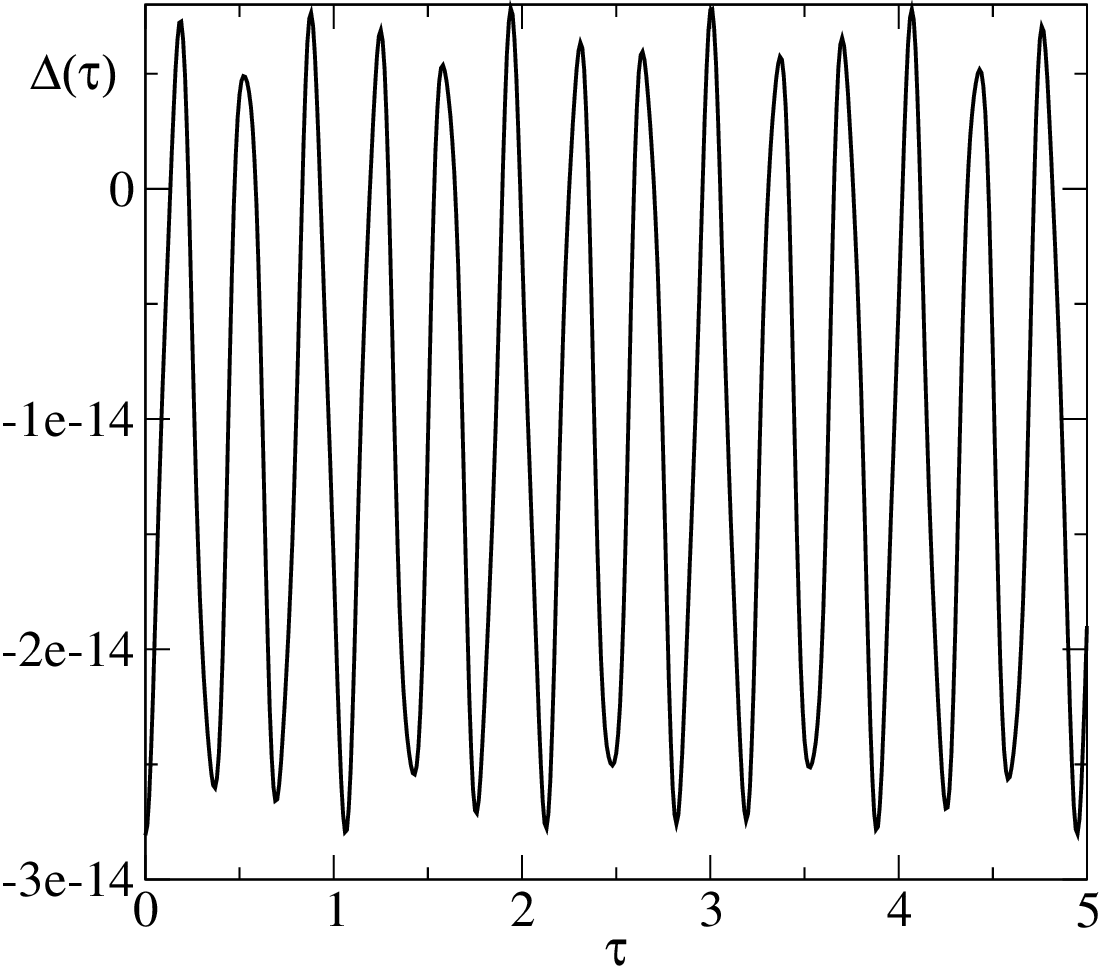}
\caption{Temperature oscillation of a gas  in circular motion around a massive object $r=7\mathcal{M}$ with Earth radius.}
\lb{fig2}
\end{figure}
%%%%%%%%%%%%%%%%%%%%%%%%%%%%%%%%%%%%%%%%%%%%%%%%%%%%%%%%%%%%%%%%%%%%%%%%%%%%%%%%%%%%%%%%%%%%%%%%%%%%%%%%%%%%%%%%%%
%%%%%%%%%%%%%%%%%%%%%%%%%%%%%%%%%%%%%%%%%%%%%%%%%%%%%%%%%%%%%%%%%%%%%%%%%%%%%%%%%%%%%%%%%%%%%%%%%%%%%%%%%%%%%%

\subsection{Compact objects}
As already mentioned in the Introduction, a potentially interesting application is the behavior of matter in the accretion disks of x-ray binaries. The observed quasiperiodic oscillations are supposed to represent the effects of matter motion in strong gravitational fields \cite{Abramowicz,Klis,Lamb,Claudio}.
While a Boltzmann gas at equilibrium certainly does not provide a realistic description of accretion disks, it might be useful nevertheless as an idealized toy model which perhaps captures at least some of the features of the real situation.
In any case, the reason for the mentioned quasiperiodic oscillations appears to be unclear so far.
The relevant frequencies are supposed to be related to test-particle frequencies. How this exactly occurs is an open problem. Very likely, hydrodynamic and/or plasma effects play a role here. Apparently, one has to find out how the individual particle motion is related to the dynamics of the medium which is described in terms of fluid or plasma quantities.
The point that our model makes is that it establishes a link between the motion of individual particles and thermodynamical quantities such as temperature and energy density. How idealized the model might ever be, it translates particle oscillations into oscillations of fluid dynamical quantities.
Such a feature should also be a necessary ingredient in more realistic models.

Tentatively, we apply our model to strong-field configurations which are typically discussed in the literature (see, e.g., Ref. \cite{Klis}).
For a neutron star with mass $M=1.4M_{\odot}$, where $M_{\odot}$ is the solar mass, the Schwarzschild radius is
$r_{S} = 2 \frac{GM}{c^{2}} \approx 4.1\ \mathrm{km}$. Let us consider a circular orbit at
$r = 5 r_{S} = 10 \mathcal{M} \approx 20.5\ \mathrm{km}$. This corresponds to  frequencies of about $\omega_{N} = 4.62 \cdot 10^{3}\ \mathrm{Hz}$.
Likewise let us consider a black hole of ten solar masses $M = 10M_{\odot}$. It has a Schwarzschild radius
$r_{S} = 2 \frac{GM}{c^{2}} \approx 29.5 \ \mathrm{km}$.
For circular orbits $r = 5 r_{S} = 10\mathcal{M} \approx 148\ \mathrm{km}$ the frequencies are of the order of $\omega_{N} = 6.4 \cdot 10^{2}\ \mathrm{Hz}$. As already mentioned, for any $r = 10 \mathcal{M}$ the frequency ratio $\Omega/\omega = \sqrt{2.5} \approx 1.58$
is of the order of the  frequently observed ratio 3/2 from x-ray binaries.
\\
\ \\
\begin{figure}
 \includegraphics[width=0.5\textwidth]{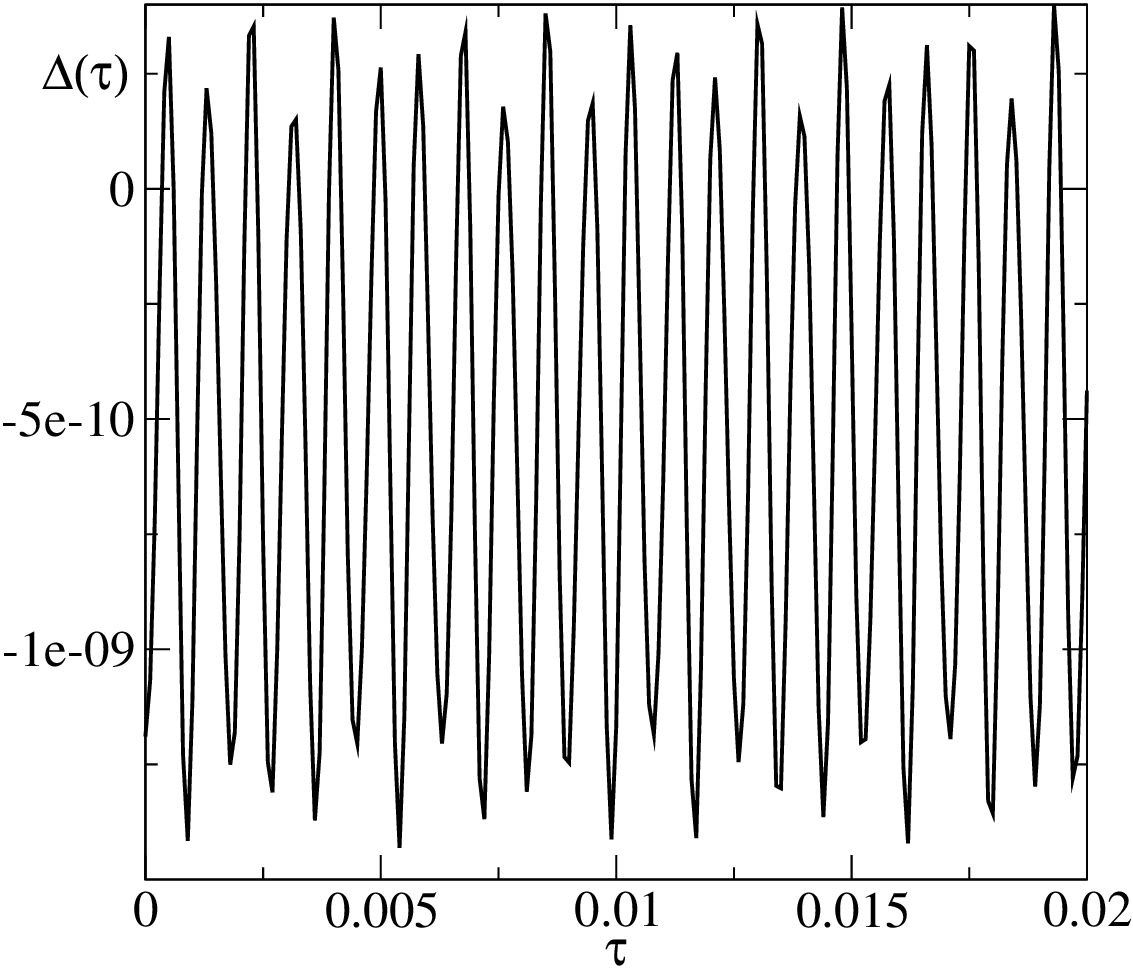}
\caption{Temperature oscillation of a gas  in circular motion around a neutron star with  $r=10\mathcal{M}$ and $M=1.4M_{\odot}$.}
\lb{fig3}
\end{figure}
\\
\ \\
\begin{figure}
 \includegraphics[width=0.5\textwidth]{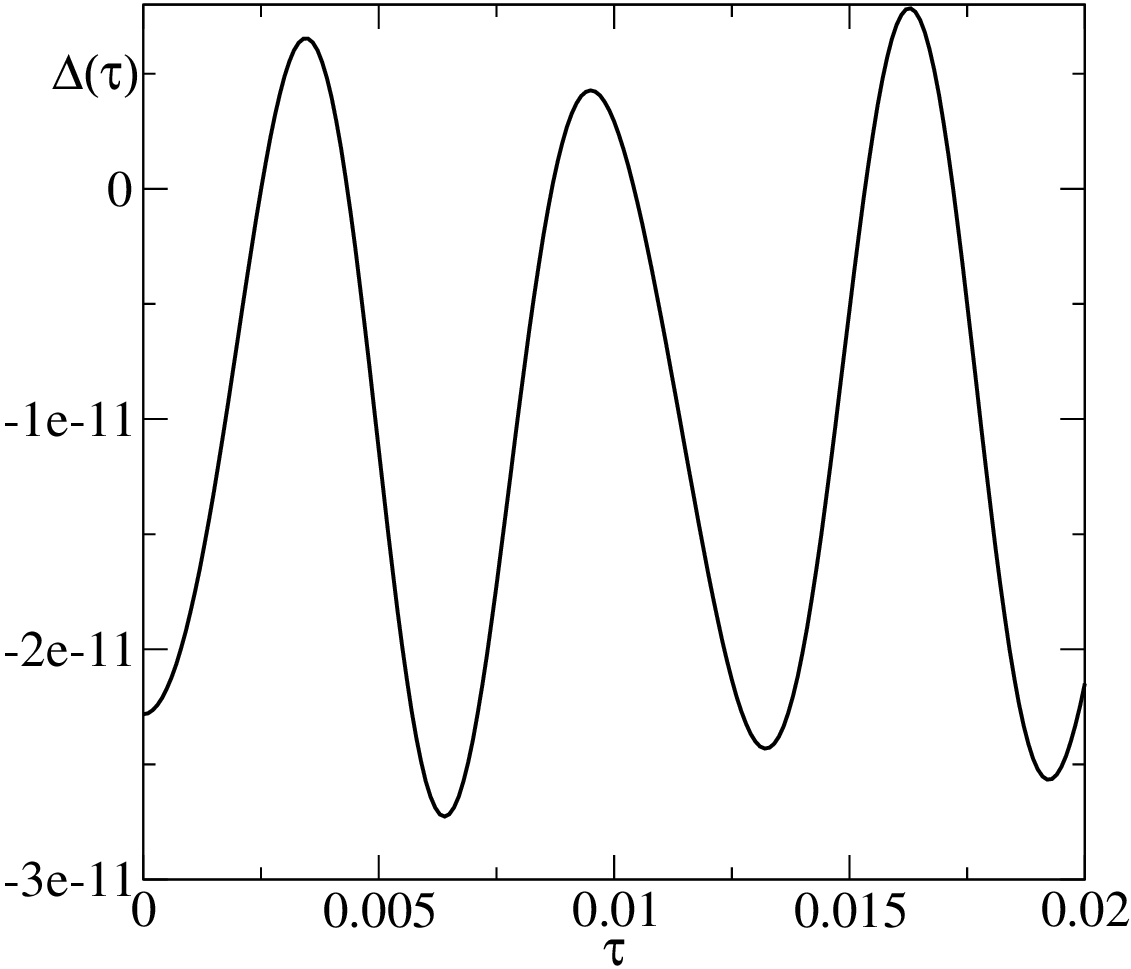}
\caption{Temperature oscillation of a gas  in circular motion around a black hole with $r=10\mathcal{M}$ and $M=10M_{\odot}$.}
\lb{fig4}
\end{figure}

Figures~\ref{fig3} and \ref{fig4} visualize the temperature oscillations of a gas in circular motion around a neutron star with $r=10\mathcal{M}$ and $M=1.4M_{\odot}$ and a black hole with $r=10\mathcal{M}$ and $M=10M_{\odot}$, respectively.
Because of the factor $\frac{\mathcal{M}}{r^{3}}$ in the expression (\ref{deltatau}) for $\Delta(\tau)$  the amplitude of the oscillations is larger for the motion in the field of the neutron star.

\section{Summary}
\label{summary}

A Boltzmann gas at equilibrium may be seen as the simplest exactly calculable  matter model that one may think of.
Although being a highly idealized configuration it sets a benchmark for more realistic models.
We have shown here that the equilibrium condition, represented by Tolman's law, dictates the entire thermohydrodynamics of the gas, including its behavior in strong gravitational fields. The temperature profile of the Boltzmann gas in circular geodesic motion in the Schwarzschild field turns out to be determined by oscillations with two frequencies, $2\omega = 2\sqrt{\frac{GM}{r^{3}}}\sqrt{\frac{r-6\mathcal{M}}{r-3\mathcal{M}}}$ and $2\Omega = 2\sqrt{\frac{GM}{r^{3}}}\sqrt{\frac{r}{r-3\mathcal{M}}}$, the difference of which is a purely general-relativistic effect.
 The oscillation frequencies of the temperature and of the other thermodynamic quantities like the energy per particle are exactly twice the frequencies for the test-particle motion. This feature is traced back to the parabolic temperature profile around the circular geodesic which, in turn, is a direct consequence of Tolman's law, applied to (modified) Fermi normal coordinates. This extends the concept of a proper reference frame to second-order deviations of the metric from the locally Minkowskian frame, carried by a comoving observer.
 Thus, the equilibrium condition allows us to relate properties of the individual particle motion to thermohydrodynamical variables.
 Different from the test-particle oscillation frequencies (\ref{ophi}) and (\ref{or}) in Schwarzschild coordinates, reviewed in Sec.~\ref{Schwarzschild}, a comoving observer does not measure a precession within the orbital plane since oscillations occur with the same frequency $2\omega$ both in radial and in tangential directions. But the oscillation frequency $2\Omega$ perpendicular to the orbital plane does not coincide with the frequency in the plane.
 The frequency difference is almost negligible in the vicinity of the planets of the Solar System.
 It may crucially affect, however, the matter dynamics close to compact astrophysical objects like neutron stars or black holes.\\
\ \\
\acknowledgments
This work  has been supported by the Conselho Nacional de Desenvolvimento Cient\'{\i}fico e
Tecnol\'ogico (CNPq), Brazil. W.Z. thanks Claudio German\`{a} for a useful discussion.

\end{document}